\documentclass[12pt]{article}
\usepackage{graphicx}
\def \bea{\begin{eqnarray}}
\def \beq{\begin{equation}}

\def \eea{\end{eqnarray}}
\def \eeq{\end{equation}}

\def \s{\sqrt{2}}

\begin{document}
\renewcommand{\thetable}{\Roman{table}}
\begin{flushright}
EFI-06-14 \\
hep-ph/0608102 \\
August 2006 \\
\end{flushright}

\centerline{\bf EFFECTS OF S-WAVE THRESHOLDS}
\bigskip
\centerline{Jonathan L. Rosner}
\medskip
\centerline{\it Enrico Fermi Institute and Department of Physics,
University of Chicago}
\centerline{\it 5640 South Ellis Avenue, Chicago, IL 60637}
\bigskip
\centerline{\large Abstract}
\bigskip
The opening of a new S-wave threshold is frequently accompanied by an
abrupt dip in the magnitude of an amplitude for an already-open channel.
One familiar example is the behavior of the $I=0$ S-wave $\pi \pi$ scattering
amplitude at $K \bar K$ threshold.  Numerous other examples of this
phenomenon in recent data are noted, and a unified description of the
underlying dynamics is sought.
\bigskip

\leftline{PACS numbers: 11.55.-m, 11.80.-m, 11.80.Gw, 13.75.-n}
\bigskip

\section{INTRODUCTION}
\bigskip

The rapid drop in the magnitude of the $I=0$ S-wave $\pi \pi$ scattering
amplitude near a center-of-mass energy $E_{\rm cm} \simeq 1$ GeV has been
known for many years.  It appears to be associated with the rapid passage
of the elastic phase shift through $180^\circ$ near the center-of-mass
energy at which the amplitude
becomes highly inelastic as a result of the opening of the $K \bar K$
threshold \cite{Flatte:1972rz}.  A resonance with $J^{PC} = 0^{++}$
($J=$ total angular momentum, $P=$ parity, $C=$ charge-conjugation
eigenvalue) now known as $f_0(980)$ \cite{PDG}, coupling both to
$\pi \pi$ and $K \bar K$, appears to be responsible for the rapid variation of
the $I=J=0$ elastic phase shift $\delta_0^0$.  For detailed discussions of
the amplitude and phase shift, see Refs.\ \cite{Au:1986vs} and
\cite{Bugg:2005nt}.

A wide variety of reactions in particle physics, many of which have been
observed only recently, display a similar rapid drop in the magnitude of an
S-wave amplitude in one channel when another channel opens up.  The present
article is devoted to a discussion of this phenomenon and its possible
dynamical implications, with suggestions for further experimental study.

The vanishing or rapid decrease of cross sections is familiar from fields
outside particle physics.  For example:

\begin{itemize}

\item The {\it Ramsauer-Townsend effect} \cite{rte} corresponds to a minimum
of the elastic cross section at which a phase shift for S-wave scattering
goes through $180^\circ$.  Writing the $S$-matrix as $S = \eta e^{2 i \delta}$,
where $\eta$ is the inelasticity parameter and $\delta$ is the phase shift,
one sees that for $\eta = 1$ and $\delta = \pi$, the scattering amplitude
$f = (S - 1)/(2 i k)$ vanishes.  (Here $k$ is the c.m.\ three-momentum.)

\item Cusps in S-wave scattering cross sections occur at thresholds for
{\it any} new channels \cite{Wigner:1948}.  This behavior has recently been
discussed in Ref.\ \cite{Bugg:2004rk} in the context of several processes to be
mentioned here.

\item {\it Monochromatic neutrons} may be produced by utilizing the vanishing
absorption cross sections of neutrons of certain energies on specific nuclei
\cite{monon}.

\end{itemize}

Within particle physics, a number of recently observed dips appear to be
correlated with S-wave thresholds:

\begin{enumerate}

\item The value of $R \equiv \sigma(e^+ e^- \to {\rm hadrons})/\sigma(e^+ e^-
\to \mu^+ \mu^-)$ drops sharply around $\sqrt{s} = 4.26$ GeV\cite{Bai:2001ct},
which happens to be just below the threshold for production of $D \bar D_1 + $
charge conjugate (c.c.), where $D$ and $D_1$ are charmed mesons with $J^{PC} =
0^-$ and $1^+$, respectively \cite{Bai:2001ct}.

\item With the advent of high-statistics Dalitz plots for heavy meson decays
one often sees dips and edges correlated with thresholds \cite{Rosner:2006ag}.
The $\pi \pi$ spectrum in $D^0 \to K_S \pi^+ \pi^-$ \cite{Aubert:2005iz,%
BeKspipi,Asner:2003uz} shows discontinuities both at $M(\omega)$
(a rapid fall-off in $M(\pi \pi)$ due to $\rho$--$\omega$ interference) and
around 1 GeV/$c^2$ (a sharp rise due to rescattering from $K \bar K$).
A recent $D^0 \to K^+ K^- \pi^0$ Dalitz plot based on CLEO data \cite{Paras06}
contains depopulated regions near $m(K^\pm \pi^0) \simeq$ 1 GeV/$c^2$ which
may be due to the opening of the $K \pi^0 \to K \eta$ S-wave threshold
or to a vanishing S-wave $K \pi$ amplitude between a low-energy $K \pi$
resonance (``$\kappa$'') and a higher resonance.  A candidate for such
a resonance exists around 1430 MeV/$c^2$ \cite{PDG}.

\item Diffractive photoproduction of $3 \pi^+ 3 \pi^-$  exhibits a dip near
$p \bar p$ threshold \cite{Frabetti:2001ah,Frabetti:2003pw}.  This dip also
occurs in the $3 \pi^+ 3 \pi^-$ spectrum produced in radiative return in
higher-energy $e^+ e^-$ collisions, i.e., in $e^+ e^- \to \gamma 3 \pi^+
3 \pi^-$, observed by the BaBar Collaboration at SLAC \cite{Coleman:2006}.

\item The $p \bar p$ spectrum produced in radiative return \cite{Coleman:2006}
exhibits dips at $\sim 2.15$ GeV/$c^2$ and $\sim 3.0$ GeV/$c^2$, which lie just
below the respective thresholds for $p \bar \Delta(1232)$ and $N(1520) \bar
N(1520)$, respectively.

\end{enumerate}

We shall discuss these and others in Section II.  A possible dynamical origin
of these effects is posed in Section III.  Implications for further experiments
are noted in Section IV, while Section V summarizes.

\section{ILLUSTRATIONS OF S-WAVE THRESHOLDS}

\subsection{Cusp in $\pi^0 \pi^0$ spectrum at $\pi^+ \pi^-$ threshold}

The $\pi^0 \pi^0$ S-wave scattering amplitude is expected to have a cusp
at $\pi^+ \pi^-$ threshold \cite{Meissner:1997fa,Meissner:1997gf}.
This behavior can be studied in the decay $K^+ \to \pi^+ \pi^0 \pi^0$, where
the contribution from the $\pi^+ \pi^+ \pi^-$ intermediate state allows one
to study the charge-exchange reaction $\pi^+ \pi^- \to \pi^0 \pi^0$ and thus
to measure the $\pi \pi$ S-wave scattering length difference $a_0-a_2$
\cite{Cabibbo:2005ez}.  The CERN NA48 Collaboration has performed such a
measurement, finding $(a_0-a_2)m_{\pi^+} = 0.264\pm 0.006~({\rm stat})
\pm 0.004~({\rm sys}) \pm 0.013~({\rm ext})$ \cite{Batley:2005ax} in
remarkable agreement with the prediction \cite{Cabibbo:2005ez}
$0.265\pm0.004$.  One can also study this effect in $\pi^+ \pi^-$ atoms.
In this manner the DIRAC Collaboration measured $|a_0-a_2| =
(0.264^{+0.033}_{-0.020})/m_{\pi^+}$ \cite{Adeva:2005pg}.

\subsection{Cusp in $\pi^0 p$ spectrum at $\pi^+ n$ threshold}

In the photoproduction reaction $\gamma p \to \pi^0 p$ the $\pi^+ n$ threshold
lies a few MeV above the $\pi^0 p$ threshold as a result of the
$\pi^+$--$\pi^0$ and $n$--$p$ mass differences.  The real part of the
$\pi^0$ photoproduction electric dipole amplitude $E_{0+}$ shows a pronounced
cusp at $\pi^+ n$ threshold \cite{Schmidt:2001vg}, in accord with predictions
of chiral pertubation theory \cite{Meissner:1997gf,MeissnerPC}. 

\subsection{Behavior of $R$ in $e^+ e^-$ annihilations}

The parameter $R$ describing the cross section for $e^+ e^- \to {\rm hadrons}$
normalized by $\sigma(e^+ e^- \to \mu^+ \mu^-)$ describes, {\it on the
average}, the sum of squares of charges $Q_i$ of those quarks which can be
produced at a given energy:
\beq
R = \sum_i Q_i^2 \left( 1 + \frac{\alpha_S}{\pi} + \ldots \right)~~~,
\eeq
where the $\alpha_S/\pi$ term is the leading QCD correction.  Thus, below charm
threshold one expects the average value of $R$ to be slightly more than 2, as
verified by a recent measurement \cite{Ablikim:2006aj} averaged between
$E_{\rm cm} = 3.650$ and 3.6648 GeV: $\bar R_{uds} = 2.224 \pm 0.019 \pm
0.089$.

\begin{figure}
\begin{center}
\includegraphics[height=4.9in]{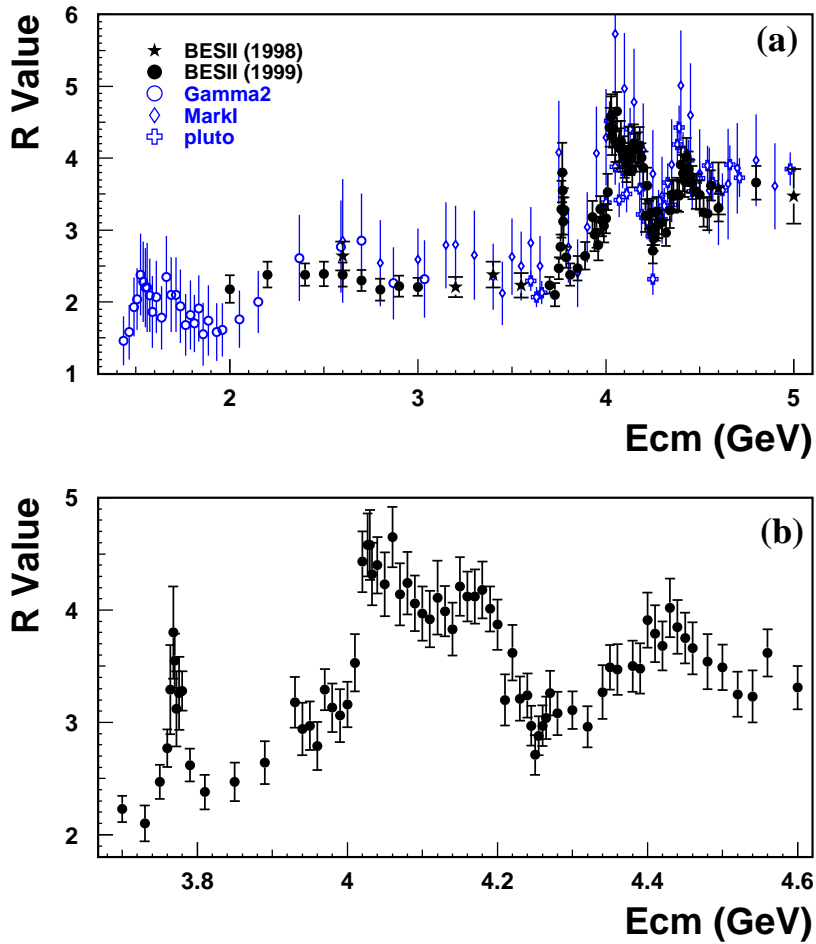}
\end{center}
\caption{(a) Compilation of $R$ values from $E_{\rm cm} = 1.4$ to 5 GeV.
(b) BES results from 3.7 to 4.6 GeV.  From Ref.\ \cite{Bai:2001ct}.
\label{fig:R}}
\end{figure}

The change in $R$ due to the opening of the charmed quark threshold should be
\beq
\Delta R_c = 3 (2/3)^2 \left( 1 + \frac{\alpha_S}{\pi} + \ldots \right)~~~,
\eeq
or slightly more than 4/3.  As one sees from Fig.\ \ref{fig:R}, this may be
true on the average, but there are strong fluctuations which are usually
ascribed to resonances around $E_{\rm cm} \simeq 4040$, 4160, and 4415 MeV
\cite{PDG}.  However, equally striking is the sharp dip in $R$,
which drops from $4.01 \pm 0.14 \pm 0.14$ at 4190 MeV to $2.71 \pm 0.12 \pm
0.13$ at 4250 MeV.  This decrease is just about the amount that would
correspond to total suppression of charm production.

The reaction $e^+ e^- \to c \bar c$ produces a $c \bar c$ system with $J^{PC} =
1^{--}$ which one expects corresponds mainly to a $^3S_1$ state.  Production of
the $^3D_1$ state, while allowed by $J$, $P$, and $C$ conservation, is
expected to be suppressed because the D-wave quarkonium wave function vanishes
at the origin \cite{Novikov:1977dq}.  The $c \bar c$ system now must fragment
into hadronic final states.

The lowest available channels are $D \bar D$ (threshold 3.73 GeV), $D \bar D^*
+ {\rm c.c.}$ (threshold 3.87 GeV), and $D^* \bar D^*$ (threshold 4.02 GeV).
Each of these corresponds to production of a charmed meson-antimeson pair in
a P-wave.  (In principle an F-wave is also possible, but highly suppressed, for
$D^* \bar D^*$.)  The cross section for production in a state of orbital
angular momentum $\ell$ grows as $(p^*)^{2 \ell + 1}$, where $p^*$ is the
magnitude of either particle's three-momentum in the center of mass.  Thus
unless a resonance is present [as is the case for $\psi(4040)$], one does
not expect abrupt effects at any of these thresholds.

The lowest-lying meson-antimeson channel with $J^{PC} = 1^{--}$ into which $c
\bar c$ can fragment with zero relative orbital angular momentum $\ell$ of the
meson-antimeson pair is $D \bar D_1 - {\rm c.c.}$ \cite{Close:2005iz}.
Here $D_1$ is a P-wave bound
state of a charmed quark and a light ($\bar u$ or $\bar d$) antiquark with $J^P
= 1^+$.  The minus sign corresponds to the negative $C$ eigenvalue.  The
lightest established candidate for $D_1$ has a mass of about 2.42 GeV/$c^2$,
corresponding to a threshold of 4.285 GeV.  It is this threshold that we
associate with the dip in $R$ between 4.19 and 4.25 GeV.

In principle there can be a lighter $J^P = 1^+$ non-strange $D$ meson $D^*_1$,
as two are expected as mixtures of $^3P_1$ and $^1P_1$ states.  If so, the
relevant threshold would lie below 4.285 GeV.  It might be smeared out, as
the $D^*_1$ is expected to be broader than the (observed) $D_1$ \cite{HQS}.
Moreover, a $^3P_0$ state $D_0$ could be sufficiently light that the lowest
S-wave meson pair with $J^{PC} = 1^{--}$ is $D^* \bar D_0 - {\rm c.c.}$ Then
its mass would have to be lower than about 2.28 GeV/$c^2$.  Again, since
a $^3P_0$ state is expected to be relatively broad, the threshold might be
broadened correspondingly.

Just above the dip in $R$ there is a newly discovered state, the $Y(4260)$,
with $J^{PC} = 1^{--}$ identified by its production in radiative return
\cite{Aubert:2005rm} and in a direct $e^+ e^-$ scan \cite{Coan:2006rv}.  We
shall discuss its possible interpretation in the next Section, but its mass
and quantum numbers suggest some relation to the dip mentioned here.
 
\subsection{Dip in $3 \pi^+ 3 \pi^-$ spectra at $p \bar p$ threshold}

\begin{figure}
\begin{center}
\includegraphics[height=5in]{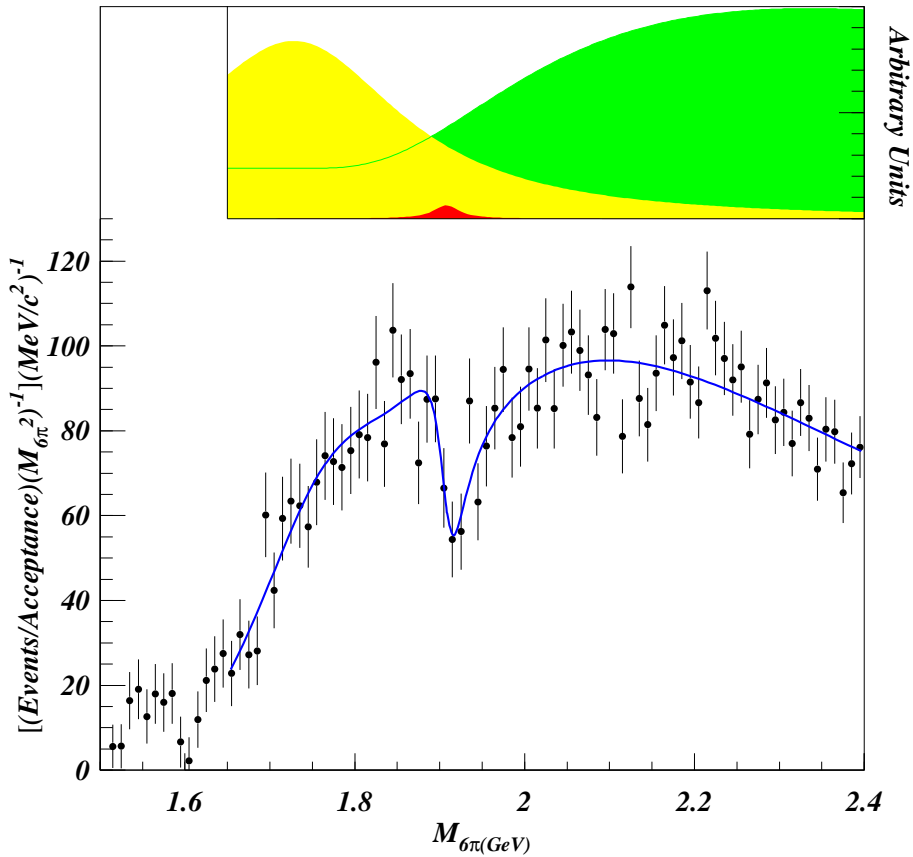}
\end{center}
\caption{Spectrum of diffractively produced $3 \pi^+ 3 \pi^-$
\cite{Frabetti:2003pw}, together with results of a fit with two resonances and
continuum.  The inset shows the relative fraction of each amplitude without
interference.
\label{fig:6pidip}}
\end{figure}

The diffractive photoproduction of $3 \pi^+ 3 \pi^-$ leads to a spectrum with a
pronounced dip near 1.9 GeV/$c^2$ \cite{Frabetti:2001ah,Frabetti:2003pw},
as shown in Fig.\ \ref{fig:6pidip}.  This
feature can be reproduced by a $1^{--}$ resonance with $M = 1.91 \pm 0.01$
GeV/$c^2$ and width $\Gamma = 37 \pm 13$ MeV interfering destructively with a
broader $1^{--}$ resonance at lower mass.  In Ref.\ \cite{Frabetti:2003pw} this
resonance is considered unlikely to be a nucleon-antinucleon bound state
because it is not seen in the reaction $\bar n p \to 3 \pi^+ 2 \pi^- \pi^0$
studied by the OBELIX Collaboration \cite{Agnello:2002kp}.  However, this
interpretation needs to be re-examined if, as noted here, the dip is really
due to the opening of a new channel, in which case a resonance pole, if
present at all, may not be located in the expected place.  This aspect will
be examined in the next Section.

The $3 \pi^+ 3 \pi^-$ spectrum also has been studied in the radiative
return reaction $e^+ e^- \to \gamma 3 \pi^+ 3 \pi^-$ by the BaBar
Collaboration \cite{Coleman:2006} in the course of a systematic examination
of all final states contributing to $R$ below about 3 GeV.  In contrast to
the photoproduction reaction, where the diffractive nature of $3 \pi^+ 3 \pi^-$
production suggests but does not firmly imply that the multi-pion system has
$J^{PC} = 1^{--}$, here the identification is unambiguous.  A dip is again
seen at 1.9 GeV/$c^2$.

\subsection{Dips in $p \bar p$ spectra at higher thresholds}

The $p \bar p$ spectrum studied in radiative return \cite{Coleman:2006} shows
two dips:  an appreciable one at 2.15 GeV/$c^2$ and a less-significant one
at 3 GeV/$c^2$.  These both can be correlated with important S-wave thresholds.

The first S-wave quasi-two-body threshold in the $J^{PC} = 1^{--}$ channel
consisting of a nucleon-antinucleon state corresponds to $N \bar \Delta -
{\rm c.c.}$, and occurs around 2.17 GeV/$c^2$.  By necessity it occurs in
the isovector channel, since $I_N = 1/2$ while $I_\Delta = 3/2$.  This
implies that the threshold effect should be equally strong in the $p \bar p$
and $n \bar n$ channels, a prediction which will be difficult to check.
However, it also implies that the production ratios for $p \bar \Delta^-$
and $n \bar \Delta^0$ should be equal, which may be verifiable using
missing-mass techniques.

The next-highest nucleon resonance above the $\Delta$ is the Roper resonance,
with a mass around $M_R \simeq 1.44$ GeV/$c^2$.  There does not seem to be
a prominent dip around $M_R + M_N \simeq 2.38$ GeV/$c^2$.  The first
{\it negative}-parity nucleon resonance is $N(1520)$, with $J^P = 3/2^-$.
If produced in a relative S-wave with another resonance in a state with
$J^{PC} = 1^{--}$, that resonance must also have negative parity, so the
corresponding threshold is 2(1.52) = 3.04 GeV/$c^2$.  Here there does seem
to be a slight dip in the $p \bar p$ spectrum.  In principle both the
isovector and isoscalar channels can be affected.  The production of
$N^0(1520) \bar N^0(1520)$ is easily detected through the $p \pi^- \bar p
\pi^+$ final state, but detection of $N^+(1520) \bar N^-(1520)$ requires
observation of $p \pi^0 \bar p \pi^0$, somewhat more challenging.

\section{MOLECULES AND BOUND STATES}

\subsection{Non-elementary nature of light scalar mesons}

The lightest scalar mesons may be denoted $f_0(600) = \sigma$, $K_0^*(800) =
\kappa$, $f_0(980)$, and $a_0(980)$ \cite{PDG}.  Although they have been
variously assigned to $q \bar q$, $qq \bar q \bar q$, and meson-meson
dynamical resonances, it seems difficult to avoid the conclusion that mesonic
(not just quark) degrees of freedom play a key role in their properties.  (See,
e.g., \cite{Bugg:2005nt,Tornqvist:1995kr,Baru:2003qq,Oller:1998zr,%
vanBeveren:2006ua}, and references therein.)  The $\sigma$ and $\kappa$ may be
dynamically generated using just current algebra, unitarity, and crossing
symmetry \cite{Oller:1998zr,BG71,Lehmann:1972kv,VB85}, as discussed in Ref.\
\cite{Rosner:2006ag}.  The $K \bar K$ thresholds play a key role in the
properties of the $f_0(980)$ and $a_0(980)$, whose masses may be strongly
affected by coupling to the $K \bar K$ channels \cite{BuggPC}.  The resonance
$f_0(980)$ below $K \bar K$ threshold has a pole not far from the real axis;
its width is 40--100 MeV \cite{PDG}.  The $f_0(980)$ drives the elastic $\pi
\pi$ phase rapidly through $180^\circ$ when its effects are combined with the
more-slowly-varying $\pi \pi$ S-wave behavior of the $\sigma$
\cite{Caprini:2005zr,Pennington:2006qi,Bugg:2006gc}.

\subsection{The $Y(4260)$ as a $D \bar D_1 - {\rm c.c.}$ bound state}

The observation of the $Y(4260)$ has sparked many interpretations.  It has
variously been identified as a conventional 4S quarkonium level
\cite{Llanes-Estrada:2005vf}, displacing the $\psi(4415)$ \cite{Barnes:2005pb}
in this role; a two-quark-two-antiquark state \cite{Maiani:2005pe}; or a hybrid
\cite{Close:2005iz,Zhu:2005hp}, corresponding to excitation of gluonic degrees
of freedom.  The favored decay in this last scenario is precisely to one
S-wave and one P-wave meson, for example $D \bar D_1 - {\rm c.c.}$  If
this channel is closed, one may still be able to observe its effects through
the off-shell decay of $D_1$ to $D^* \pi$.

If the closed $D \bar D_1 - {\rm c.c.}$ channel is responsible for the
$Y(4260)$, one might expect a closed $D_s \bar D_{s1} - {\rm c.c.}$ channel to
generate similar behavior at higher energy \cite{ClosePC}.  The threshold
for this channel is about 4430 MeV$/c^2$, so the $\psi(4415)$ might have
enhanced coupling to it or to $D_s^* \bar D_{s0}$, which has a similar
threshold \cite{Barnes:2005pb}.

\subsection{Effect of a $^3S_1$ $p \bar p$ bound state on $\gamma^{(*)} \to
3 \pi^+ 3 \pi^-$}

The six-pion channel with isospin $I = I_3 = 1$ does not display resonant
activity above $\bar n p$ threshold \cite{Agnello:2002kp}.  However, if there
is a state with strong coupling to $3 \pi^+ 3 \pi^-$ and $p \bar p$
below $p \bar p$ threshold, one may expect behavior similar to what is
observed in the $\pi \pi$ S-wave channel as described above.  [Note added:
a satisfactory fit to the behavior illustrated in Fig.\ 2 was obtained in
Ref.\ \cite{Bugg:2004rk} from a cusp effect alone, without recourse to a
resonance.  One objection to a $\bar p p$ resonance is the absence of
a peak near threshold in the cross section for $\bar p p \to \bar n n$.]

The $^1S_0$ $p \bar p$ system exhibits behavior
characteristic of a shallowly-bound state with mass around 1835 MeV/$c^2$.
Such a state is consistent with what is observed in the radiative decays
$J/\psi \to \gamma p \bar p$ \cite{Bai:2003sw} and $J/\psi \to \gamma \eta'
\pi^+ \pi^-$ \cite{Ablikim:2005um}.  Here, if the state is produced in the
expected way via conversion of two gluons emitted in the decay $J/\psi \to
\gamma g g$, it is likely to be isoscalar.

\subsection{A state near $\omega \phi$ and $K^* \bar K^*$ threshold}

The BES II Collaboration has observed an enhancement in $M(\omega \phi)$
near threshold in the decay $J/\psi \to \gamma \omega \phi$
\cite{Ablikim:2006dw}.  The state emerges from a partial wave analysis as a
candidate for $J^P = 0^+$ with mass $M = 1812^{+19}_{-26} \pm 18$ MeV/$c^2$
and width $\Gamma = 105 \pm 20 \pm 28$ MeV/$c^2$.  The spin and parity are
consistent with being an S-wave state of $\omega \phi$ just about 10 MeV/$c^2$
above threshold.  It is interesting that the $K^* \bar K^*$ and $\omega \phi$
thresholds are only about 10--20 MeV/$c^2$ apart depending on the $K^*$
charges.  We suspect the reaction $\omega \phi \leftrightarrow K^* \bar K^*$
plays an important part in stabilizing this resonance.  One interpretation
of this state \cite{Li:2006ru} implies a larger partial width to $K^* \bar
K^*$ than to $\omega \phi$.

\subsection{$p \bar \Delta - {\rm c.c.}$ and $N(1520) \bar N(1520)$ bound
states}

The relative parities of a $p$ and a $\bar \Delta$ are negative.  The S-wave
bound states of a suitably antisymmetrized state will have $P=C=-$. The allowed
$J$ values for $p \bar \Delta$ are 1 and 2, but only $J=1$ couples
to the virtual photon.

The relative parities of $N(1520)$ and $\bar N(1520)$ also are negative, so all
their S-wave bound states have negative parity.  The total spins $S$ can
take on all values between 0 and 3, but one expects (as for spin-1/2 particles)
that the charge-conjugation eigenvalue of the pair is $C=(-1)^{L+S}$. The $J=1$
state thus is permitted by $C$ to couple to the virtual photon.

The dynamics which would give rise to bound states in the $J=1$ channels are
unclear.  By analogy with the $p \bar p$ system mentioned
above one might expect bound states with more than one $J$ value.

\subsection{The $\Lambda(1405)$ as a $\bar K N$ bound state}

Another system which has been known for nearly 50 years to have features in
common with the bound states mentioned above is the $I=0$ S-wave kaon-nucleon
system, which has a $J^P = 1/2^-$ resonance at 1405 MeV/$c^2$, about $26 \pm 4$
MeV/$c^2$ below $K^- p$ threshold \cite{Dalitz:1959dn}.  This state, the
$\Lambda(1405)$, decays essentially 100\% of the time to $\Sigma \pi$. Although
it may be viewed in the quark model as a $uds$ state with orbital
angular momentum $\ell = 1$, the detailed properties of the $\Lambda(1405)$
can be viewed equally successfully in terms of the final-state channels to
which it couples \cite{Petersen:1972qk,Faiman:1976gg,Isgur:1978xi}.  This
is true of a number of other members of the negative-parity baryon
70-dimensional multiplet of SU(6).  For example, the $N(1535)$, also with
$J^P = 1/2^-$, can be viewed both as an $L = 1$ three-quark state and as
a dynamical effect with large coupling to $\eta N$ \cite{Tuan}.

\subsection{Excited $\Xi$ as a $\Sigma \bar K$ threshold effect}

The BaBar Collaboration has recently analyzed the mass, width, and spin of a
baryon with strangeness $-2$ near 1685 MeV/$c^2$ \cite{Aubert:2006ux},
observed in the decay $\Lambda_c \to \Lambda \bar K^0 K^+$.  The existence
of this state, and its correlation with $\Sigma \bar K$ threshold, has been
known for a number of years (see the references in \cite{PDG}), but the
BaBar analysis finds its spin to be consistent with 1/2, which would
correspond to an S-wave $\bar K \Sigma$ state.  The mass, found in Ref.\
\cite{Aubert:2006ux} to be $1684.7 \pm 1.3^{+2.2}_{-1.6}$ MeV/$c^2$, probably
lies above the $K^- \Sigma^+$ threshold of 1683.0 MeV/$c^2$ but below the
$\bar K^0 \Sigma^0$ threshold of 1690.3 MeV/$c^2$.  Thus in some sense
one could regard it as a $\bar K^0 \Sigma^0$ S-wave bound state.

\subsection{Peak in $M(\Lambda p)$ at $\Sigma N$ threshold}

For a number of years, it has been known that the $\Lambda p$ mass spectrum
in $K^- d \to \Lambda p \pi^-$ exhibits a sharp peak around $\Sigma N$
threshold \cite{Jaffe}.  Early references with the largest data samples include
Refs.\ \cite{Tan:1969jq} ($p_K \simeq 0$) and \cite{Braun:1977ma} ($p_K = 0.7$
GeV/$c$);~
others may be found in \cite{PDG84}.  The mass of this state is consistent
with being equal to $M(\Sigma^+ n) = 2128.93$ MeV/$c^2$; the $\Sigma^0
p$ threshold at 2130.91 MeV/$c^2$ is 2 MeV/$c^2$ higher.  This state has been
interpreted as the $^3S_1$ $S=-1,~I=1/2$ partner of the deuteron in the
$\overline{10}$-dimensional representation of SU(3)
\cite{Oakes:1963,Dalitz:1980zc}.  A recent discussion \cite{Bugg:2004rk},
however, finds that it is difficult to determine whether the data demand an
actual pole in addition to a threshold cusp.  This ambiguity is common to a
number of examples we have discussed here.

\subsection{The $X(3872)$ as a $D^0 \bar D^{*0} + {\rm c.c.}$ bound state}

The suggestion that charmed mesons might form molecules or bound states with
one another was made shortly after their discovery \cite{Voloshin:1976ap}.
It now appears that one has a good candidate for such a state.
The $X(3872)$, discovered by Belle in $B$ decays \cite{Choi:2003ue}
and confirmed by BaBar \cite{Aubert:2004ns} and in hadronic production
\cite{Acosta:2003zx}, decays predominantly into $J/\psi \pi^+ \pi^-$.
It has many features in common with an S-wave bound state of $(D^0 \bar D^{*0}
+ \bar D^0 D^{*0})/ \sqrt{2} \sim c \bar c u \bar u$ with $J^{PC} = 1^{++}$
\cite{Close:2003sg}.  Its $J^{PC} = 1^{++}$ assignment is supported by its
recently reported observation in the $D^0 \bar D^0 \pi^0$ channel
\cite{Gokhroo:2006bt}.  It has recently been described more generally as being
associated with a large scattering length in $D^0$--$\bar D^{*0}$ scattering
\cite{Bugg:2004rk,Braaten:2005jj}.

\subsection{The $D_{s0}(2317)$ and $D^*_{s1}(2460)$ as
$D^{(*)}K$ bound states}

The lowest-lying $J^P = 0^+$ and $1^+$ $c \bar s$ mesons \cite{Aubert:2003fg}
have turned out to be lighter than predicted in most \cite{Godfrey:1985xj,%
Cahn:2003cw} (but not all \cite{Matsuki:1997da}) quarkonium calculations.
In fact, they lie below their respective $D K$ and $D^* K$ thresholds, and thus
must decay via emission of a photon or an isospin-violating $\pi^0$ to a lower
$c \bar s$ state.  While low masses are predicted \cite{Bardeen:2003kt} by
viewing these states as parity-doublets of the $D_s(0^-)$ and $D^*_s(1^-)$
$c \bar s$ ground states, one can also view them as bound states of $D^{(*)}K$
\cite{Barnes:2003dj,vanBeveren:2003kd,Guo:2006fu} (the binding energy in each
case would be 41 MeV), or as $c \bar s$ states with masses lowered by coupling
to $D^{(*)}K$ channels \cite{Close:2004ip}.

\subsection{Is there a rule for bound state formation?}

The feature which the above examples have in common is the coexistence of at
least two channels, one closed and one open, near the energy at which either a
dip or a peak is observed.  This is a necessary but probably not sufficient
condition for what is termed a {\it Feshbach resonance} \cite{Feshbach:1958nx},
which has been used to great advantage in the study of Bose-Einstein
condensates \cite{BE}.  Such a resonance is characterized by a phase shift
increasing rapidly by $180^\circ$ as the energy rises through the resonance.
This is how the $I=0$ S-wave $\pi \pi$ phase shift behaves \cite{Au:1986vs},
but it is not known whether the phase shifts in the other channels we have
mentioned behave similarly.  However, all of these channels are dominated by a
single partial wave (e.g., $^3S_1$ for the $c \bar c$ pair produced directly or
via radiative return in $e^+ e^-$ collisions), so the possibility of a
rapidly decreasing or vanishing amplitude exists.  The importance of
hadronic as well as quark degrees of freedom in processes such as those we have
discussed has been stressed in Refs.\ \cite{Bugg:2005nt,Bugg:2004rk,%
Tornqvist:1995kr,Baru:2003qq,Oller:1998zr,vanBeveren:2006ua,Close:2004ip,%
Close:2003tv,vanBeveren:2002zt}.

A regularity governing resonance formation was noted some time ago
\cite{Rosner:1973fq}.  If two mesons are allowed by the quark model to
resonate, they do so for $p^*< p_0^{MM} = 350$ MeV/$c$, where $p^*$ is the
c.m. momentum. The corresponding value for meson-baryon systems is
$p_0^{MB} = 250$ MeV/$c$.  In order to form non-exotic resonances, an
antiquark in a meson must annihilate with the corresponding quark in the other
meson or baryon.

The $\omega \phi$ threshold of Ref.\ \cite{Ablikim:2006dw} is one possible
counterexample, because the quark flavors in the two decay products are
distinct from one another:  $(u \bar u + d \bar d)/\s$ for the $\omega$ and
$s \bar s$ for the $\phi$.  However, if this resonance couples strongly
not only to $\omega \phi$ but also to $K^* \bar K^*$, the latter component
may be chiefly responsible through the proposed quark-antiquark annihilation
mechanism for the resonant behavior. 

It is possible that a stronger form of the above rule holds when neither
resonating particle is a pion.  The pion may be considered as anomalously light
in view of its role as a Nambu-Goldstone boson of spontaneously broken chiral
SU(2) $\times$ SU(2).  Thus, $\pi^+$ and $\pi^-$ form a $\rho$ meson (or
possibly a $\sigma$) well above threshold, but $K$ and $\bar K$ seem to exhibit
their first resonances in the $I=0$ state $f_0(980)$ and $I=1$ state
$a_0(980)$, both of which lie below threshold.  The $K^-$ and $p$ form the
$\Lambda(1405)$, also below threshold.  This behavior is frequent enough that a
more general dynamical principle may be at work.

\section{IMPLICATIONS FOR EXPERIMENTS}

\subsection{Dip in $R$: hadronic makeup of final states}

If the dip in $R$ is due to a new threshold in the hadronization of $c \bar c$,
it should be confined to final states consisting of charmed mesons and
charmed antimesons (with possible additional pions).  The cross section for
$e^+ e^-$ production of non-charmed final states should not be affected.

\subsection{Isovector states produced via $B \to \bar D^{(*)} W^{*+}$ decay}

The dip in the $p \bar p$ spectrum at $p \bar \Delta^-$ threshold, if indeed
due to this threshold, must be occurring in the isovector channel.  Then
production of $p \bar n$ by the charged weak current should exhibit a
similar dip at $p \bar \Delta^0$ threshold.  One should be able to observe this
effect in the decay (e.g.) $B^0 \to D^{*-} p \bar n$, where the $\bar n$ is
reconstructed via missing mass.  Similarly, the decay $B^0 \to D^{*-}
(6 \pi)^+$ should show a dip in the six-pion effective mass spectrum at
$p \bar n$ threshold, or around 1.9 GeV/$c^2$.

\subsection{Elastic $\Sigma \pi$ scattering at $\Lambda(1405)$}

The SELEX Collaboration has completed a program of studies with a $\Sigma^-$
beam \cite{Russ:2000bs}.  Although the main focus of this experiment was
charm production, it was capable of studying $\Sigma^- \pi^+$ scattering
through the peripheral process $\Sigma^- p \to X n$ with a cut on small
momentum transfer to isolate pion exchange.  If the $K^- p$ threshold plays
a role similar to that noted in the above examples, one would expect a sharp
dip in the cross sections for $\Sigma^- \pi^+ \to \Sigma^\mp \pi^\pm$ in
the vicinity of $\Lambda(1405)$.

\subsection{Threshold $K \eta$ or sub-threshold $K \eta'$ resonance}

The depletion of the $K \pi$ spectrum just above 1 GeV/$c^2$ in the Dalitz
plot for $D^0 \to K^+ K^- \pi^0$ \cite{Paras06} occurs just around the $K \eta$
threshold.  This depletion should be confirmed by the analysis of a larger
$D^0 \to K^+ K^- \pi^0$ data sample \cite{Aubert:2006xn} and the $D^0 \to K^+
K^- \eta$ Dalitz plot studied to determine if there is an enhancement at
$K \eta$ threshold.  However, the coupling of a $0^+$ resonance to $K \eta$ is
expected to be suppressed \cite{Lipkin:1980tk}; coupling to $K \eta'$ is
favored.  An alternative interpretation is that the dip above 1 GeV/$c^2$
is due to the vanishing of the S-wave amplitude between a low-energy
$J^P = 0^+$ $K \pi$ resonance known as the $\kappa$ (for a discussion, see
Refs.\ \cite{Bugg:2005nt,Bugg:2005xx}) and a higher $0^+$ $K \pi$ resonance
(e.g., the $K_0^*(1430)$ \cite{PDG}), presumably with large $K \eta'$
coupling \cite{Lipkin:1980tk}.

\subsection{Bound states of $\bar B^{(*)} K$}

If the $D_{s0}(2317)$ and $D^*_{s1}(2460)$ are, respectively, states of
$D K$ and $D^* K$ each bound by 41 MeV, perhaps by annihilation of a $u$ or $d$
quark in the $K$ with the corresponding antiquark in the $D^{(*)}$, one might
expect a $J^P = 0^+$ $b \bar s$ state with a mass of about $M(B) + M(K) - 41
\simeq 5733$ MeV/$c^2$ and a $J^P = 1^+$ $b \bar s$ state with a mass of about
$M(B^*) + M(K) - 41 \simeq 5778$ MeV/$c^2$.  (Here we have taken the average
of charged and neutral kaon masses, and ignored changes in hyperfine energies
when replacing $c$ by $b$.  Calculations in Ref.\ \cite{Guo:2006fu} find a
$\bar B K$ bound state at $5725 \pm 39$ MeV/$c^2$ and a $\bar B^* K$ bound
state at $5778 \pm 7$ MeV/$c^2$, while the most recent of
Refs.\ \cite{Matsuki:1997da} finds a $\bar B K$ bound state at 5627 MeV/$c^2$
and two $\bar B^* K$ bound states at 5660 and 5781 MeV/$c^2$.) In analogy with
the charmed-strange system, these states might be expected to decay via photon
or $\pi^0$ emission to $B_s$ and/or $B^*_s$, subject to the usual spin-parity
selection rules.

\subsection{Effects in exotic baryon-antibaryon channels}

The regularity noted above for meson-meson and meson-baryon resonance
formation, that when an antiquark in a meson can annihilate a quark in a meson
or baryon then one expects a resonance to be formed below a certain value of
$p^* < p_0$, was generalized to baryon-antibaryon resonances
\cite{Rosner:1973fq} to predict a corresponding value $p_0^{B \bar B} = 200$
MeV/$c$.  This rule, if correct, would predict the existence of exotic
baryon-antibaryon resonances \cite{Rosner:1968si}, which have never been
seen, not far above baryon-antibaryon threshold.  Examples of manifestly exotic
channels are $\Delta^{++} \bar n$ and $\bar \Lambda p \pi^+$, with minimal
quark content $uu \bar d \bar d$ and $u u \bar d \bar s$, respectively.
Some recent suggestions for observing such resonances in $B$ decays were made
in Ref.\ \cite{Rosner:2003ia}.  

A stronger version of the above resonance formation model is suggested in
the present article:  If neither incident particle is a pion, at least one
resonance may be formed {\it below threshold}.  Thus, the search for
exotic mesons may require one to study spectra well below baryon-antibaryon
threshold, where ground-state exotic mesons are indeed expected
\cite{Jaffe:1976ig}.

\section{SUMMARY}

The rapid decrease of the S-wave $I=0$ $\pi \pi$ scattering amplitude near
$K \bar K$ threshold is seen to be mirrored in a host of other phenomena,
whose origin may reflect similar physics.  In the case of the coupled
$\pi \pi$ and $K \bar K$ channels, a key role is played by the $f_0(980)$
resonance, whose presence is responsible for a rapid increase of the
elastic $I=J=0$ phase shift $\delta^0_0$ through $180^\circ$, where the
S-wave $\pi \pi$ amplitude vanishes just before becoming highly inelastic.
Other effects which may be regarded within the same framework include the
dip in $R$ for $e^+ e^- \to$ (hadrons) around a center-of-mass energy
of 4.25 GeV and several dips in the cross section for photoproduction or
$e^+ e^-$ production of specific final states near the thresholds for others.

It is suggested that all these effects may be associated with the formation
of bound states in the channels which are about to open, with a possible
relation to Feshbach resonances which similarly occur in coupled-channel
problems.  Tests of this proposal include the observation of a dip in $I=0$
S-wave $\Sigma \pi$ scattering near $K^- p$ threshold and the presence of
exotic baryon-antibaryon resonances below threshold.  This phenomenon does
not appear to be universal, but widespread.  Further dynamical information
remains to be found in order to predict its occurrence reliably.

\section*{ACKNOWLEDGMENTS}

I thank E. van Beveren, D. Bugg, F. Close, J. Collar, F.-K. Guo, C. Hanhart, R.
Jaffe, T. Matsuki, U. Mei{\ss}ner, J. Pel\'aez, G. Rupp, and S. F. Tuan for
helpful comments, and the Newman Laboratory of
Elementary-Particle Physics, Cornell University, for its kind hospitality
during part of this investigation.  This work was supported in part by the
United States Department of Energy under Grant No.\ DE FG02 90ER40560.

\end{document}